\documentclass[12pt]{article}

\usepackage{amsfonts} %paquetes de símbolos
\usepackage{amsmath}
\usepackage{amsthm}
\usepackage[usenames,dvipsnames]{color} %cores
\usepackage[english]{babel}

\usepackage{amssymb}

\usepackage{bm}
\usepackage[inner=3cm,outer=3cm,bottom=3cm,top=3cm]{geometry}
\usepackage{latexsym,pdfsync,xcolor,graphicx}
\usepackage{array}
\usepackage{subfig}
\usepackage{tikz}
\usepackage{natbib}
\usepackage{appendix}

\usepackage{hyperref}

\usepackage{subfig}
\usepackage{enumerate} 
\usepackage{multirow} 
\usepackage{rotating}
\usepackage{float}
\usepackage{textcomp}
\usepackage{booktabs}
\usepackage{lscape}
\usepackage{titling}
\usepackage[ruled,vlined]{algorithm2e}
\SetKwInput{KwInput}{Input}                % Set the Input
\SetKwInput{KwOutput}{Output}              % set the Output

\numberwithin{equation}{section}
\theoremstyle{plain}

\newtheorem{prop}{Proposition}
\newtheorem{lemma}{Lemma}

\theoremstyle{remark}

\DeclareMathOperator*{\argmax}{arg\,max}

\begin{document}

\title{\large Nonparametric multimodal regression for circular data} 

\author{\small Mar\'ia Alonso-Pena and Rosa M. Crujeiras}
\date{}

%\address[labelusc]{}
%\address[labelkul]{Department of Mathematics, KU Leuven}

\maketitle
\vspace{-1.8cm}
\footnotesize
\begin{center}
Department of Statistics, Mathematical Analysis and Optimization, Universidade de \\Santiago de Compostela.
\end{center}
\vspace{0.3cm}

\hrule
\vspace{-0.4cm}
\normalsize
\noindent
\begin{center}
\textbf{Abstract}
\end{center}
\vspace{-0.3cm}
Multimodal regression estimation methods are introduced for regression models involving circular response and/or covariate. The regression estimators are based on the maximization of the conditional densities of the response variable over the covariate. Conditional versions of the mean shift and the circular mean shift algorithms are used to obtain the regression estimators. The asymptotic properties of the estimators are studied and the problem of bandwidth selection is discussed.

\vspace{0.1cm}
 
\vspace{0.1cm}

\hrule

\section{Introduction}
\label{sec:intro}

There is a diverse range of practical situations where one may encounter random variables which are not defined on Euclidean spaces. It is the case of angles, directions, events or periodic observations, which can be thought as data on the unit circumference (circular data). Circular data can be found in many different fields: biology (orientation of the red wood ants in reaction to different stimuli; \citet{Jander1957}), geology (cross-beds; \citet{SenGupta_Rao1966}),  environmetrics and oceanography (wind and waves directions; \citet{Jona-Lasinio_etal2012} and \citet{Oliveira_etal_2014}), medicine (sudden infant death syndrome; \citet{Mooney_etal2003}) or ecology (wild fire occurrences in the Iberian Peninsula; \citet{Ameijeiras-Alonso_etal2019b}).
The particular nature of this kind of observations calls for the need of specific modeling and inferential methods beyond those tailored for data on the real line. \citet{Jammalamadaka_SenGupta_2001} and \citet{Pewsey_etal2013} present a detailed review of circular statistics. Circular data can also be viewed as a particular case of directional data, defined in a hypersphere of arbitrary dimension \citep{Mardia_Jupp_2000,Ley_Verdebout2017}. Hence, methods for hyperspherical data can be adapted to the circumference.

\begin{figure}[t]
	\centering
	\subfloat{
		\includegraphics[width=0.28\textwidth]{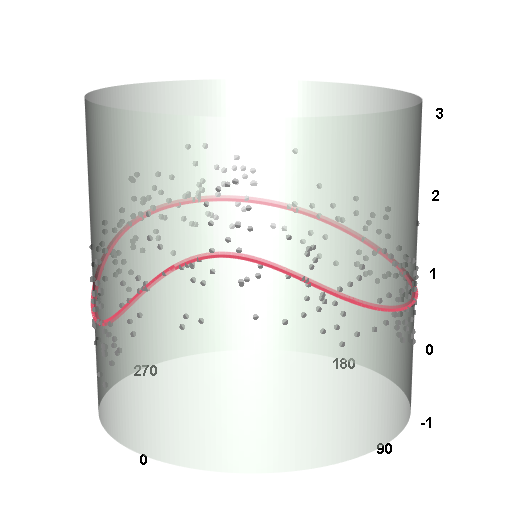}
		\label{fig:example_mean_regression_circlin}}
	\hfill
	\subfloat{
		\includegraphics[width=0.33\textwidth]{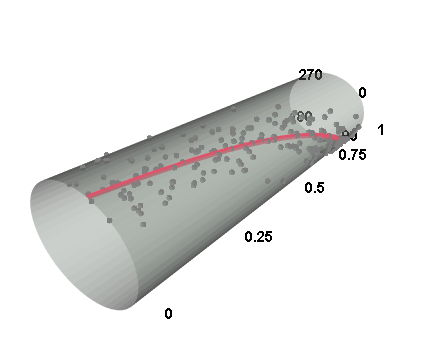}
		\label{fig:example_mean_regression_lincirc}}
	\hfill
	\subfloat{
		\includegraphics[width=0.28\textwidth]{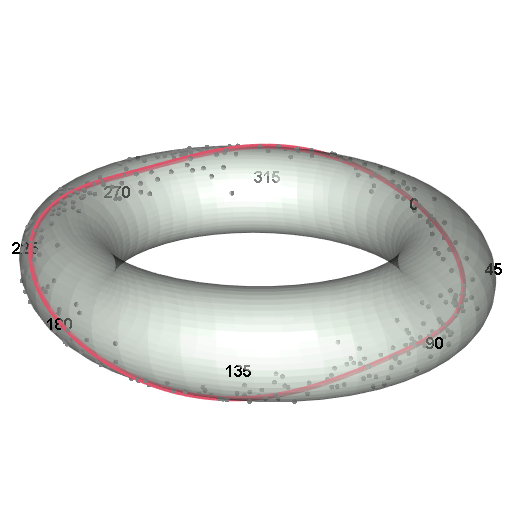}
		\label{fig:example_mean_regression_circcirc}} 
	
	\bigskip % some vertical space between subfigures a/b and c/d...
	\subfloat{
	\includegraphics[width=0.28\textwidth]{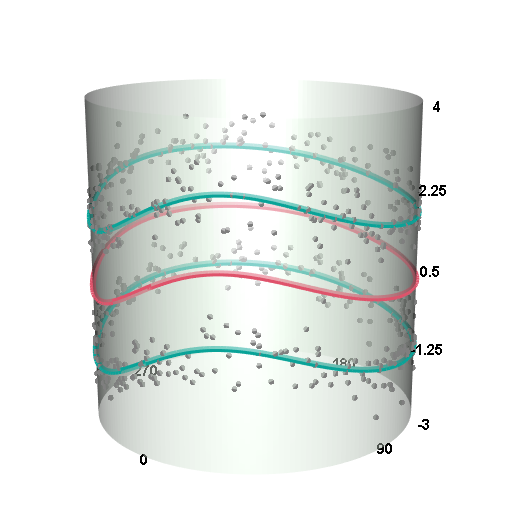}
	\label{fig:example_mode_regression_circlin}}
\hfill
\subfloat{
	\includegraphics[width=0.33\textwidth]{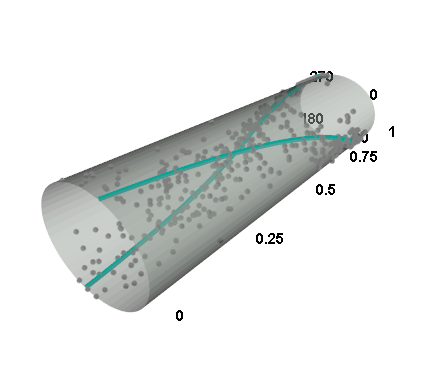}
	\label{fig:example_mode_regression_lincirc}}
\hfill
\subfloat{
	\includegraphics[width=0.28\textwidth]{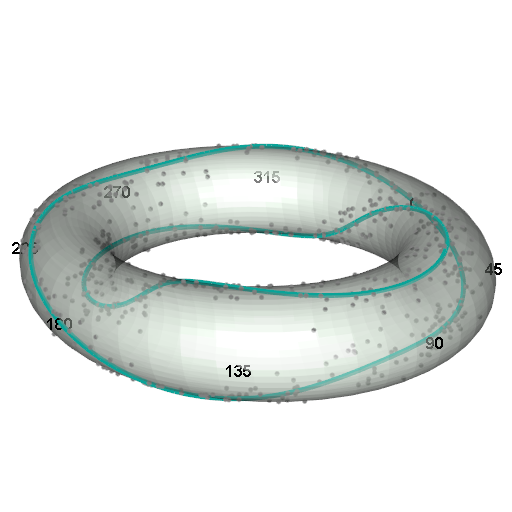}
	\label{fig:example_mode_regression_circcirc}} 
	
	\caption{Representations on the cylinder and on the torus of simulated data from regression models with a circular predictor and a real-valued response (left column), a real-valued predictor and a circular response (middle column) and a circular predictor and a circular response (right column). Red curves represent the conditional means in each scenario. In the bottom row the conditional modes are also depicted in green.}
	\label{fig:examples_mean_mode}
\end{figure}

Circular measurements may be accompanied by other observations, either defined on the unit circumference or on the real line. In such cases it can be of interest to model the relationship between the variables from a regression perspective. Three different contexts, involving circular variables, can be considered: circular predictor and real-valued response, real-valued predictor and circular response and both circular predictor and response. The regression functions on the first two scenarios can be represented on the cylinder, while the third one may be visualized in the torus, as displayed in the top row of Figure~\ref{fig:examples_mean_mode}. Several proposals for parametric regression models in the different contexts can be found in \citet{Jammalamadaka_SenGupta_2001}. However, parametric models may not be sufficiently flexible to model more complex relationships between variables, so nonparametric methods present a useful alternative. For instance, \citet{DiMarzio_etal2009} consider a local-polynomial regression method for circular predictors and a real-valued response by including periodic kernels. For the cases where the response is a circular variable, \citet{DiMarzio_etal2012} propose a kernel-type regression estimator by smoothing the cosine and sine components of the response. 

Although the aforementioned parametric and nonparametic methods have different formulation contexts, they all consider the conditional mean as the target function to estimate.  However, the classical approach of \textit{regression to the mean} might not be suitable in cases where the conditional density is highly skewed or multimodal. The bottom row in Figure~\ref{fig:examples_mean_mode} presents simulated data from circular regression models where the conditional densities of the response over the covariate are bimodal. It can be observed that the conditional mean function is not a good representative of the relationships between the variables. In fact, the mean regression function is not even defined on the bottom middle and bottom right examples of Figure~\ref{fig:examples_mean_mode}, as the conditional densities are bimodal and symmetric. Instead, the conditional local modes provide a better alternative to model the data, leading to the so-called \textit{multimodal regression} approach in which, instead of a function, the target is a multifunction or multi-valued function. 

The idea of estimating the local modes of the conditional density to approach regression in the Euclidean context was first introduced by \citet{Scott1992}. \citet{Einbeck_Tutz2006} propose using a conditional version of the mean shift algorithm \citep{Fukunaga_Hostetler1975,Cheng1995,Comaniciu_Meer2002} to estimate the regression multifunction. The theoretical properties of this approach were studied by \citet{Chen_etal2016} and a recent review of multimodal regression for Euclidean variables can be found in \citet{Chen2018}. The mean shift algorithm was extended to directional variables by \citet{Oba_etal2005} in the context of nonparametric modal clustering and it was also studied by \citet{Kobayashi_Otsu_2010}. The convergence of the algorithm and some other theoretical results were recently obtained by \citet{Zhang_Chen2020}.

The goal of this manuscript is to introduce the use of the conditional versions of the Euclidean and circular mean shift to estimate the modal regression multifunction when the response and/or the covariate have a circular nature. Section~\ref{sec:modal_reg_circ} presents the regression scenarios and details the algorithms for estimating the circular regression multifunctions. Some theoretical results are given in Section~\ref{sec:theoretical_results}, while the problem of bandwidth selection is discussed in Section~\ref{sec:bandwidth}.

\section{On the use of mean shift for circular \\ modal regression}
\label{sec:modal_reg_circ}

Nonparametric multimodal regression aims for estimating the conditional density of a response over a covariate with a flexible nonparametric estimator and, afterwards, compute the conditional local modes with the so-called mean shift algorithm. In this section we will detail the algorithms for multimodal regression multifunctions in the different circular contexts depicted in Figure~\ref{fig:examples_mean_mode}. We will distinguish the case where the response variable is a real-valued variable (left panels in Figure~\ref{fig:examples_mean_mode}), where we make use of the usual conditional mean shift algorithm, and the cases where the response is circular (center and right panels in Figure~\ref{fig:examples_mean_mode}), where the circular conditional mean shift is employed.

\subsection{Real-valued response} \label{subsec:real-valued_response}
In this section, we will consider a circular predictor, $\Theta$, with support on the circumference of the unit circle, $\mathbb{T}=(-\pi,\pi]$ and a real-valued response, denoted by $Y$, with support on $\Omega\subset\mathbb{R}$. Let $\{(\Theta_j,Y_j)\}_{j=1}^n$ be a bivariate sample from $(\Theta,Y)$.

In order to model the relationship  between the predictor and the response, we will consider the modal regression multifunction, which for each value $\theta\in \mathbb{T}$ is defined as the set of local modes of the  conditional density function $f(y|\theta)$:
$$ M(\theta)= \bigg \{ m_t(\theta):\  t=1,...,p(\theta)  \bigg \}, $$ 
where $p(\theta)$ is the number of branches of the multifunction at  $\Theta=\theta$ and $m_t(\theta)$, $t=1,...,p(\theta)$, are the local modes, that is:
$$ m_t(\theta)=\argmax_{a \in S_t} f(a|\theta), $$
with $S_t$ being a closed interval ($S_t\subset\Omega$) and where the maximum is taken in the interior of $S_t$. Therefore, for a fixed $\theta\in \mathbb{T}$, the local modes (i.e. local maxima) of the conditional density function are given by
\begin{equation}
M(\theta)=\bigg \{ y: \frac{\partial}{\partial y}f(y|\theta)=0,\frac{\partial^2}{\partial y^2}f(y|\theta)<0\bigg\}.
\label{eq:circlin_multifunction}
\end{equation}
The estimation of the set $M(\theta)$ is carried out through an indirect approach: first, the conditional density is estimated and afterwards the conditional local modes are computed. For the estimation of the conditional density of the response over the covariate, we will use a kernel type estimator. We will first review the kernel density estimators on the real line and on the circumference, and subsequently present the kernel type estimator of the conditional density function.

% Change: start by explaining KDE (real-valued) --> KDE (circular)
% --> conditional KDE 
Start by considering the response variable, $Y$, and its density function, which with an abuse of notation will be denoted by $f(y)$. Given the sample of the response values, $\{Y_j\}_{j=1}^n$, the kernel density estimator of $f$ \citep{Parzen1962,Rosenblatt1956} is given by
\begin{equation}
	\hat{f}(y)=\frac{1}{n}\sum_{j=1}^{n}L_h(y-Y_j),
	\label{eq:KDE}
\end{equation}
where $L_h$ is a \textit{linear} kernel (\textit{i.e.}, a symmetric around zero density) with bandwidth $h$:
$$ L_h(y-Y_j)=\frac{1}{h}L\left(\frac{y-Y_j}{h}\right). $$
Usually, the Gaussian density is used as the kernel in practice. The selection of the smoothing parameter, $h$, greatly affects the performance of (\ref{eq:KDE}), since if $h$ is too small the estimator $\hat{f}$ will be undersmoothed. On the contrary, if $h$ is too large, an oversmoothed estimator will be obtained. See \citet{WandJones1995} for details on estimator (\ref{eq:KDE}).

Consider now the circular random variable $\Theta$ and the sample $\{\Theta_j\}_{j=1}^n$. The density function of $\Theta$ can be estimated with the circular kernel density estimator \citep{Fisher1989}, which is obtained as
\begin{equation}
	\hat{f}(\theta)=\frac{1}{n}\sum_{j=1}^{n}K_\kappa(\theta-\Theta_j)
	\label{eq:KDE_circ}
\end{equation}
where $K_\kappa$ is a circular kernel (a circular density function symmetric about the zero direction) with concentration parameter $\kappa$. It must be noted that $\kappa$ plays a role opposite to $h$ in the smoothing of the kernel density estimator: if $\kappa$ is small (low concentration), an oversmoothed estimator is obtained and, if the value of $\kappa$ is large (high concentration), the kernel density estimator will be undersmooothed. A kernel usually employed in practice is the von Mises, with density function 
$$ f(\theta;\mu,\kappa)=\frac{1}{I_0(\kappa)}\exp\{\kappa \cos(\theta - \mu)\},$$
where $I_0(\kappa)$ denotes the modified Bessel function of the first kind and order zero.

We now consider the problem of estimating the density of $Y$ given the value of $\Theta$ with a kernel type estimator. We estimate the conditional density $f(y|\theta)$ as
\begin{equation}
	\hat{f}(y|\theta)=\dfrac{\sum_{j=1}^{n}K_\kappa(\theta-\Theta_j)L_h(y-Y_j)}{\sum_{j=1}^{n}K_\kappa(\theta-\Theta_j)}.
	\label{eq:cond_density_circlin}
\end{equation}
Note that the numerator in (\ref{eq:cond_density_circlin}) is proportional to the product kernel estimator of the joint density in the cylinder \citep{GarciaPortugues_etal2013} and the denominator is proportional to the circular kernel density estimator in (\ref{eq:KDE_circ}). Thus, the kernel density estimator of the conditional density depends on two smoothing parameters, $\kappa$ and $h$. The role of the smoothing parameters will be discussed in Section~\ref{sec:bandwidth}.

In order to estimate the regression multifunction, we plug the expression in (\ref{eq:cond_density_circlin}) into (\ref{eq:circlin_multifunction}) obtaining
\begin{equation}
\hat{M}(\theta)=\bigg \{ y\in \Omega : \frac{\partial}{\partial y }\hat{f}(y|\theta)=0, \frac{\partial^2}{\partial y^2 }\hat{f}(y|\theta)<0 \bigg \}.
\label{eq:estimator_mutifunction_circlin}
\end{equation}
The computation of the conditional local modes is not straightforward and, thus, the conditional mean shift algorithm is used. We will assume $L$ is a radially symmetric kernel with profile $l$, so that $L(\cdot)=c_L\times l((\cdot)^2)$, with $c_L\in\mathbb{R}^+$. A local maximum of $\hat{f}(y|\theta)$ must satisfy
$$ \frac{\partial}{\partial y}\hat{f}(y|\theta)= \dfrac{\frac{2c_L}{h^3}\sum_{j=1}^nK_\kappa\left( \theta-\Theta_j \right)l' \bigg \{\left( \frac{y-Y_j}{h} \right)^2\bigg \}(y-Y_j)}{\sum_{j=1}^{n}K_\kappa(\theta-\Theta_j)}=0.$$
By taking $g(\cdot)=-l(\cdot)'$, we have
\begin{equation}
	\frac{\partial}{\partial y}\hat{f}(y|\theta)= \frac{-2c_L}{nh^3\hat{f}(\theta)}\sum_{j=1}^nK_\kappa\left( \theta-\Theta_j \right)g \bigg \{\left( \frac{y-Y_j}{h} \right)^2\bigg \}(y-Y_j)=0,
	\label{eq:shifts_euclidean}
\end{equation}
where $\hat{f}(\theta)$ is defined as in (\ref{eq:KDE_circ}). Note that the last factor in the sum of the first equality in  (\ref{eq:shifts_euclidean}), $(y-Y_j)$, is the shift of each datum $Y_j$ to the point $y$. Therefore, the partial derivative in (\ref{eq:shifts_euclidean}) is a weighted sum of the shifts to the point $y$. Now, by denoting $G(\cdot)=c_gg\{(\cdot)^2\}$, we have
$$ y_m=\dfrac{\sum_{j=1}^{n}K_\kappa(\theta-\Theta_j)G\left(\frac{y_m-Y_j}{h}\right)Y_j}{\sum_{j=1}^{n}K_\kappa(\theta-\Theta_j)G\left(\frac{y_m-Y_j}{h}\right)},  $$
However, we are not able to solve for $y_m$ since $y_m$ also appears in the right term of the previous equation, i.e., we have a fixed-point equation
$$ y_m=\omega(y_m), \quad \text{where} \quad\omega(\cdot)=\dfrac{\sum_{j=1}^{n}K_\kappa(\theta-\Theta_j)G\left(\frac{\cdot-Y_j}{h}\right)Y_j}{\sum_{j=1}^{n}K_\kappa(\theta-\Theta_j)G\left(\frac{\cdot-Y_j}{h}\right)}. $$
Note that when $L(\cdot)$ is a Gaussian kernel, $G(\cdot)$ is also a Gaussian kernel and the function $\omega(\cdot)$ is actually a weighted mean of the observations with weights depending on the proximity of the data to the point $(\theta,y)$. In order to compute the conditional local modes we apply the mean shift algorithm, which was proposed by \citet{Fukunaga_Hostetler1975} in the context of gradient density estimation and subsequently studied by \citet{Cheng1995,Comaniciu_Meer2002,Li_etal2007}. The convergence of the algorithm was also investigated by \citet{AliyariGhassabeh_etal2013,AliyariGhassabeh_2013,AliyariGhassabeh_2015}. The algorithm was generalized to a conditional version by \citet{Einbeck_Tutz2006} and \citet{Chen_etal2014} considered general weights in the context of mode and ridge estimation. For each value of $\theta$, we consider the so-called mean shift function, defined as 
\begin{equation}
	m(y)=\omega(y)-y, 
	\label{eq:meanshift_function}
\end{equation}
which represents the shift we get when moving from $y$ to the weighted mean of the observations with weights depending on $y$, point at which the function is evaluated. The key feature of the mean shift algorithm is that for a local maximum of the conditional density, the mean shift function takes the value zero.

Now we apply the conditional version of the algorithm with the weights being circular kernel functions, which is described in Algorithm~\ref{alg:cond_meanshift}. It is easily seen that the conditional mean shift algorithm is actually a gradient ascent method on $\hat{f}(y|\theta)$ for a fixed $\theta$, where the step size is implicitly chosen, since $m(y)$ in (\ref{eq:meanshift_function}) is proportional to and has the same direction as $\frac{\partial}{\partial y }\hat{f}(y|\theta)$.

\begin{algorithm}[ht]
	
	\SetAlgoLined
	\KwData{Sample $\{(\Theta_j,Y_j)\}_{j=1}^n$, smoothing parameters $\kappa$ and $h$.}
	
	1. Initialize mesh points $\mathcal{T}\subset \mathbb{T}$.\\
	2. For each $\theta \in \mathcal{T}$, select starting points $y_0^{(1)}(\theta),...,y_0^{(p)}(\theta)$.\\
	3. For $k=1,...,p$ iterate until convergence: $$ y_{l+1}^{(k)}=\dfrac{\sum_{j=1}^{n}K_\kappa(\theta-\Theta_j)G\left(\frac{y_l^{(k)}-Y_j}{h}\right)Y_j}{\sum_{i=1}^{n}K_\kappa(\theta-\Theta_j)G\left(\frac{y_l^{(k)}-Y_j}{h}\right)}, \quad \mbox{with} \  \ l=0,1,... $$

	\caption{Conditional mean shift }
	\label{alg:cond_meanshift}
\end{algorithm}

%\textcolor{red}{mention how to select the starting points and also that the number of starting points might be different for each $\theta$}
Note that the number of starting points in Algorithm~\ref{alg:cond_meanshift}, $p$, may be different for each value of $\theta$. \citet{Chen_etal2016} take the whole sample as the starting points, as it is usually done in nonparametric modal clustering, but we recommend a \textit{local initialization} where, for each value of $\theta$, the starting values are the sample responses closer to $\theta$.

\subsection{Circular response} \label{subsec:circular_response}

Consider now a circular response variable, $\Phi$, with support on $\mathbb{T}$ and either a real-valued predictor $X$ with support on $\Omega\subset\mathbb{R}$ or a circular predictor $\Theta$ with support on $\mathbb{T}$. We will use $\Delta$ to denote a generic predictor, with support either on the real line or on the unit circumference. Let $\{(\Delta_j,\Phi_j)\}_{j=1}^{n}$ be a random sample from $(\Delta,\Phi)$. Similarly to Section~\ref{subsec:real-valued_response}, for each value of the predictor variable, $\delta$, the modal regression multifunction is defined as
\begin{equation}
M(\delta)= \bigg \{ \phi : \frac{\partial}{\partial \phi}f(\phi|\delta)=0,\frac{\partial^2}{\partial \phi^2}f(\phi|\delta)<0 \bigg   \},
\label{eq:Multifunction_circ_resp}
\end{equation}
with $f(\phi|\delta)$ being the conditional density of $\Phi$ given the value of $\Delta$. We 4take the same indirect approach as in the previous section: estimate the conditional density and then compute its local maxima for each value of the predictor. We consider the kernel density estimator of the conditional density for a circular response \citep{DiMarzio_etal2016}. If the predictor variable is real-valued ($\Delta=X$), the estimator is given by
\begin{equation*}
\hat{f}(\phi|x)=\dfrac{\sum_{j=1}^{n}L_h\left(x-X_j\right)K_\kappa\left(\phi-\Phi_j\right)}{\sum_{j=1}^{n}L_h\left(x-X_j\right)}
\end{equation*}
In this case $L_h$ is a \textit{linear} kernel with smoothing parameter $h$ and $K_\kappa$ is a circular kernel with concentration parameter $\kappa$. If the predictor is circular ($\Delta=\Phi$), the conditional density is estimated as
\begin{equation*}
\hat{f}(\phi|\theta)=\dfrac{\sum_{j=1}^{n}K_\nu\left(\theta-\Theta_j\right)K_\kappa\left(\phi-\Phi_j\right)}{\sum_{j=1}^{n}K_\nu\left(\theta-\Theta_j\right)},
\end{equation*}
where the expression in the numerator is proportional to an estimation of the joint density with a product of circular kernels \citep{DiMarzio_etal2011}. Here, the kernel associated to the predictor has concentration parameter $\nu$ and the kernel associated to $\Phi$ has concentration $\kappa$. From now on we will denote the weights corresponding to the predictor variable ($X$ or $\Theta$) at the point $\delta$ as $w_\delta(\Delta_j),$ $j=1,...,n$. Note that such weights depend on the smoothing parameter $h$ or $\nu$ (depending on $\Delta$ being circular or scalar). Thus, 
\begin{equation*}
\hat{f}(\phi|\delta)=\frac{1}{n\hat{f}(\delta)}\sum_{j=1}^{n}w_\delta(\Delta_j)K_\kappa(\phi-\Phi_j), \quad \hat{f}(\delta)=\frac{1}{n}\sum_{j=1}^{n}w_\delta(\Delta_j).
\end{equation*} 
Consequently, the estimator of the multimodal regression function (\ref{eq:Multifunction_circ_resp}) is
\begin{equation}
\hat{M}(\delta)= \bigg \{ \phi : \frac{\partial}{\partial \phi}\hat{f}(\phi|\delta)=0,\frac{\partial^2}{\partial \phi^2}\hat{f}(\phi|\delta)<0 \bigg   \}. 
\label{eq:circ_resp_estim_Multifunction}
\end{equation}
We will assume that the circular kernel associated to the response variable satisfies the condition
\begin{equation}
K_\kappa(\cdot)=c_\kappa K[\kappa (1-\cos(\cdot))],
\label{eq:condition_circular_kernel}
\end{equation}
where $c_\kappa$ is a normalizing constant depending on $\kappa$. This matches the definition of directional kernel given in \citet{Bai_etal1988} for the particular case of circular data. In order to obtain the local maxima of $\hat{f}(\delta|\phi)$ we establish the necessary condition for a critical point, which is
$$ \frac{\partial}{\partial \phi}\hat{f}(\phi|\delta)=0. $$
Thus, by applying (\ref{eq:condition_circular_kernel}), we get
\begin{equation}
\frac{\partial}{\partial \phi}\hat{f}(\phi|\delta)  =\frac{\kappa c_\kappa}{n\hat{f}(\delta)}\sum_{j=1}^{n}w_\delta(\Delta_j)K'[\kappa(1- \cos(\phi-\Phi_j))] \sin(\phi-\Phi_j).
\label{eq:derivative1}
\end{equation}
Consequently, the derivative of the estimated conditional density with respect to $\phi$ is a weighted sum of the sines of the differences from each observation to the point $\phi$. In this case, the sine function is used for measuring the shifts from the observations to $\phi$. This is very intuitive since, if $\phi=\Phi_j$, then $\sin(\phi-\Phi_j)=0$. In addition, if the difference $\phi-\Phi_j$ is very small, then $\sin(\phi-\Phi_j)\approx \phi-\Phi_j$. Therefore, the right side of equation (\ref{eq:derivative1}) can be interpreted as a weighted sum of the shifts from the observations to $\phi$. By expanding the last factor in the right side of (\ref{eq:derivative1}) we get
\begin{equation*}
\frac{\partial}{\partial \phi}\hat{f}(\phi|\delta)  =\frac{\kappa c_\kappa}{n\hat{f}(\delta)}\sum_{j=1}^{n}w_\delta(\Delta_j)K'[\kappa(1- \cos(\phi-\Phi_j))] (\sin \phi \cos \Phi_j - \cos \phi \sin \Phi_j),
\label{eq:derivative2}
\end{equation*}
and by equating it to zero, we have
\begin{equation*}
\sin \phi \sum_{j=1}^{n}w_\delta(\Delta_j) T(\phi-\Phi_j) \cos \Phi_j =
 \cos \phi \sum_{j=1}^{n}w_\delta(\Delta_j) T(\phi-\Phi_j)\sin \Phi_j,
%\label{eq:derivative}
\end{equation*}
where $T(\cdot)=c_T K'[\kappa (1-\cos(\cdot))]$. Therefore, if we denote
\begin{equation}
S_\delta(\phi)=\sum_{j=1}^{n}w_\delta(\Delta_j) T(\phi-\Phi_j)\sin \Phi_j
\label{eq:sine}
\end{equation}
and
\begin{equation}
C_\delta(\phi)=\sum_{j=1}^{n}w_\delta(\Delta_j)  T(\phi-\Phi_j)\cos \Phi_j,
\label{eq:cosine}
\end{equation} 
we have that if $S_\delta(\phi)\neq0$ or $C_\delta(\phi)\neq0$,
$$ \phi= \mbox{atan}2\left( S_\delta(\phi), C_\delta(\phi)\right), $$
where the operator $\mbox{atan}2(a,b)$ returns the angle between the $x$-axis and the vector from the origin to $(b,a)$ \citep[see][Chap.~1]{Jammalamadaka_SenGupta_2001}. Since the quantity $S_\delta(\phi)$ is a weighted sum of $\sin \Phi_j$  and $C_\delta(\phi)$ is a weighted sum of $\cos \Phi_j$ (where the weights depend on the point ($\delta,\phi$)), we obtain that the mode estimator $\phi_m\equiv \phi_m(\delta)$ is given by
$$ \phi_m=\tilde{\omega}(\phi_m)=\mbox{atan}2\left( S_\delta(\phi_m), C_\delta(\phi_m) \right). $$
Note that the function $\tilde{\omega}(\phi)$ actually returns a weighted circular mean direction of the observations. Since in the previous expression we have a fixed-point equation, we use a mean shift-type algorithm in order to obtain the conditional mode estimator. We define the circular mean shift function as
$$ \tilde{m}(\phi)=\sin(\tilde{\omega}(\phi)-\phi). $$
As discussed before, given that for small values of $\tilde{\omega}(\phi)-\phi$ we have $\sin(\tilde{\omega}(\phi)-\phi)\approx \tilde{\omega}(\phi)-\phi$, the sine function is used to measure the shift from $\phi$ to $\tilde{\omega}(\phi)$. In addition, for a local mode of the conditional density estimator, the circular mean shift function takes the value zero. Consequently, the estimated regression multifunction (\ref{eq:circ_resp_estim_Multifunction}) is obtained by using the conditional circular mean shift algorithm, which is described in Algorithm~\ref{alg:circ_cond_meanshift}. As in Algorithm~\ref{alg:cond_meanshift}, the number of starting points, $p$, can be different for each value of $\delta$, and we recommend a local initialization.

\begin{algorithm}[ht]
	
	\SetAlgoLined
	\KwData{Sample $\{(\Delta_j,\Phi_j\}_{j=1}^n$, smoothing parameters $\kappa$ and $h$/$\nu$.}
	
	1. Initialize mesh points $\mathcal{S}\subset \Omega$ if $\Delta=X$ or $\mathcal{T}\subset \mathbb{T}$ if $\Delta=\Theta$.\\
	2. For each $\delta \in \mathcal{S}$ (or $\delta \in \mathcal{T}$), select starting points $\phi_0^{(1)}(\delta),...,\phi_0^{(p)}(\delta)$.\\
	3. For $k=1,...,p$ iterate until convergence: $$ \phi_{l+1}^{(k)}=\mbox{atan}2\left(\sum_{j=1}^nw_\delta(\Delta_j)T(\phi_{l}^{(k)}-\Phi_j)\sin\Phi_j, \sum_{j=1}^nw_\delta(\Delta_j)T(\phi_{l}^{(k)}-\Phi_j)\cos\Phi_j\right), $$
	with $l=0,1,...$

	\caption{Circular conditional mean shift }
	\label{alg:circ_cond_meanshift}
\end{algorithm}

Note that, although the derivations of the algorithms are different, the unconditional version of Algorithm~\ref{alg:circ_cond_meanshift} actually coincides with the directional mean shift proposed by \citet{Oba_etal2005} and studied by \citet{Kobayashi_Otsu_2010} and \citet{Zhang_Chen2020} in the particular case of data in the unit circumference. Consequently, the convergence result in \citet{Zhang_Chen2020} guarantees the convergence of Algorithm~\ref{alg:circ_cond_meanshift}.

\begin{figure}[ht]
	\centering
	\includegraphics[width=0.65\textwidth]{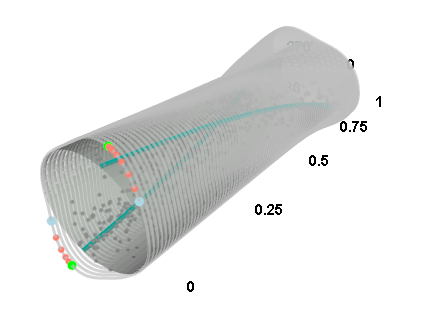}
	\caption{Representation on the cylinder of simulated data from a bimodal regression multifunction with the true regression multifunction along with the conditional density estimator and a visualization of the conditional circular mean shift algorithm. }
	\label{fig:cylinder_meanshift}
\end{figure}

Brief calculations show that the circular mean shift function $\tilde{m}(\phi) $ is proportional to and has the same direction as $\frac{\partial}{\partial \phi}\hat{f}(\phi|\delta)$. Thus, the circular mean shift can be interpreted as a circular gradient ascent method on $\hat{f}(\phi|\delta)$ (see \citet{Bonnabel_2015} and \citet{Zhang_Sra2016} for works on general gradient ascent algorithms on Riemannian manifolds). Moreover, the ascending property of the directional mean shift is proven in \citet{Kobayashi_Otsu_2010} by assuming the convexity of the kernel. This interpretation can be graphically observed in Figure~\ref{fig:cylinder_meanshift}, where Algorithm~\ref{alg:circ_cond_meanshift}, for a fixed $x\in\mathbb{R}$, is initialized at two starting points (blue dots) and converges at the two conditional local modes of the estimated density (green dots).

\section{Some theoretical results}
\label{sec:theoretical_results}

The aim of this section is to provide asymptotic error rates of the multimodal regression estimators introduced in Section~\ref{sec:modal_reg_circ}.  Since the estimation of the regression multifunctions is related to the estimation of the derivatives of the conditional density estimator, we first need to stablish consistency results for the derivatives of the conditional density functions. In fact, it is enough to consider the partial derivatives of the joint density estimator, since the regression multifunction (\ref{eq:circlin_multifunction}) is equivalent to
\begin{equation*}
M(\theta)=\bigg \{ y: \frac{\partial}{\partial y}f(\theta,y)=0,\frac{\partial^2}{\partial y^2}f(\theta,y)<0\bigg\}
\end{equation*}
and the multifunction in (\ref{eq:Multifunction_circ_resp}) is equivalent to
\begin{equation*}
M(\delta)= \bigg \{ \phi : \frac{\partial}{\partial \phi}f(\delta,\phi)=0,\frac{\partial^2}{\partial \phi^2}f(\delta,\phi)<0 \bigg   \}.
\end{equation*}

\subsection{Consistency of the joint density estimators and their \\derivatives}

Let $\Theta$ and $\Phi$ be circular random variables and let $Y$ be a real-valued random variable. Let $\hat{f}(\theta,y)$ denote the kernel density estimator of the cylindrical density, $f(\theta,y)$, given by
$$\hat{f}(\theta,y)=\frac{1}{n}\sum_{i=1}^{n}K_\kappa(\theta-\Theta_i)L_h(y-Y_i)$$
and $\hat{f}(\theta,\phi)$ denote the kernel density estimator of the joint density in the torus, 
$$\hat{f}(\theta,\phi)=\frac{1}{n}\sum_{i=1}^{n}K_\nu(\theta-\Theta_i)K_\kappa(\phi-\Phi).$$
We consider the following assumptions.
\begin{itemize}
	\item \textbf{(A1)} The joint density function is at least three times continuously differentiable and its partial derivatives are square integrable (either on $\mathbb{R}$ or on $\mathbb{T}$, depending on the nature of the variables).
	\item \textbf{(A2)} The linear kernel $L$  is a symmetric around zero linear density with finite second order moment.
	\item \textbf{(A3)} The circular kernel $K_\kappa$ satisfies $K_\kappa(\cdot)=c_\kappa K[\kappa (1-\cos(\cdot))]$ where $c_\kappa$ is a normalization constant, $K:[0,\infty)\rightarrow [0,\infty)$ and
	$$ 0<\int_{0}^{\infty}K(s)^ks^{-1/2}ds<\infty \quad 0<\int_{0}^{\infty}K^{(r)}(s)^2s^{-1/2}ds<\infty   $$
	for $k=1,2$, $r=1,2$ and
	$$ \int_{\varepsilon^{-1}}^{\infty}|K'(s)|s^{1/2}ds=O(\varepsilon). $$
	
	\item \textbf{(A4)} The smoothing parameters $h=h_n$, $\kappa=\kappa_n$ and $\nu=\nu_n$ are sequences of positive numbers such that $h_n\rightarrow0$, $\kappa_n\rightarrow \infty$, $\nu_n\rightarrow \infty$, $nh^{1+2r}\kappa^{-1/2}\rightarrow \infty$, $nh\kappa^{\frac{-(1+2r)}{2}}\rightarrow \infty$,  $n\nu^{-1/2}\kappa^{\frac{-(1+2r)}{2}}\rightarrow \infty$ as $n\rightarrow\infty$, for $r=0,1,2$.
\end{itemize} 
Condition \textbf{(A1)} is a standard smoothing condition on the population density required in order to obtain the convergence rates of the partial derivatives of the joint density estimators. Condition \textbf{(A2)} is also a usual condition on kernel density estimation literature and is satisfied by many kernels in practice, such as the normal kernel. The third condition, \textbf{(A3)}, is required by \citet{Zhang_Chen2020} in order to obtain the asymptotic convergence rates of the derivatives of the kernel density estimator in the unit hypersphere and, as noted by the authors, integrating by parts shows that the von Mises kernel satisfies this condition. The last condition, \textbf{(A4)}, is a \textit{usual} condition on the smoothing parameters. 

\begin{lemma}\label{lem:consistency_joint}
	Under conditions \textbf{(A1)}, \textbf{(A2)}, \textbf{(A3)} and \textbf{(A4)}, for fixed $x,y\in\mathbb{R}$, $\theta,\phi\in\mathbb{T}$, we have 

	$$\frac{\partial^r}{\partial y^r}\hat{f}(\theta,y)=O(\kappa^{-1}+h^2) + O_P\left(\sqrt{\frac{\kappa^{1/2}}{nh^{1+2r}}}\right),$$
	$$\frac{\partial^r}{\partial \theta^r}\hat{f}(\theta,y)=O(\kappa^{-1}+h^2) + O_P\left(\sqrt{\frac{\kappa^{(1+2r)/2}}{nh}}\right)$$
	and
	 $$\frac{\partial^r}{\partial \phi^r}\hat{f}(\theta,\phi)=O(\kappa^{-1}+\nu^{-1}) + O_P\left(\sqrt{\frac{\kappa^{(1+2r)/2}\nu^{1/2}}{n}}\right)$$
	for $r=0,1,2$.
\end{lemma}
The proof of Lemma~\ref{lem:consistency_joint} is straightforward by using the same approach of Propositions 1 and 2 in \citet{GarciaPortugues_etal2013} and Theorem 2 in \citet{Zhang_Chen2020}. Stronger consistency results for the derivatives of the joint kernel density estimator are needed in order to study the asymptotic properties of the multimodal regression estimators. We define the following function classes:
$$ \mathcal{L}=\bigg \{ v\rightarrowtail L^{(\alpha)}\left(\frac{x-v}{h}\right): x\in\mathbb{R}, h>0, \alpha=0,1,2 \bigg \}, $$
$$ \mathcal{K}=\bigg \{ \gamma \rightarrowtail D^{[\bm{\tau}]}K[\kappa (1-\cos(\theta-\gamma))]: \theta\in\mathbb{T}, \kappa>0, [[\bm{\tau}]]=0,1,2 \bigg \}, $$
where $D^{[\bm{\tau}]}=\frac{\partial^{\tau_1}}{\partial(\cos\theta)^{\tau_1}} \frac{\partial^{\tau_2}}{\partial(\sin\theta)^{\tau_2}}$ and $[[\bm{\tau}]]=\tau_1+\tau_2$.
We assume the following conditions:
\begin{itemize}
	\item \textbf{(K1)} The kernel $L$ and its two first derivatives are bounded in absolute value by $C_L$ and the collection $\mathcal{L}$ is a VC-type class, i.e., there exist $A,q>0$ such that for any $0<\varepsilon<1$, 
	$$\sup_QN(\mathcal{L},L_2(Q),C_L\varepsilon)\leq \left(\frac{A}{\varepsilon}\right)^q, $$
	where $N(T,d,\varepsilon)$ is the $\varepsilon$-covering number for a semimetric space $(T,d)$ and $Q$ is any probability measure.
	\item \textbf{(K2)} $D^{[\bm{\tau}]}K$ is bounded in absolute value by $C_K$ for $[[\bm{\tau}]]=0,1,2$ and the collection $\mathcal{K}$ is a VC-type class: there exist $A', q' >0$ such that for any $0<\varepsilon<1$,
		$$\sup_QN(\mathcal{K},L_2(Q),C_K\varepsilon)\leq \left(\frac{A'}{\varepsilon}\right)^{q'}. $$
\end{itemize} 
Conditions \textbf{(K1)} and \textbf{(K2)} correspond to standard conditions on the kernels in order to obtain the uniform convergence rates by the methods in \citet{Gine_Guillou_2002} and \citet{Einmahl_Mason2005}. Such conditions can be satisfied in practice by, for example, the Gaussian kernel in the linear case and the von Mises kernel in the circular case.

\begin{lemma}\label{lem:uniform}
		Under conditions \textbf{(A1)}, \textbf{(A2)}, \textbf{(A3)}, \textbf{(K1)} and \textbf{(K2)}, we have $$ \sup_{\theta,y}\bigg|\bigg|\frac{\partial^r}{\partial y^r}\hat{f}(\theta,y)-\frac{\partial^r}{\partial y^r}f(\theta,y)\bigg|\bigg|= O(\kappa^{-1}+h^2) + O_P\left(\sqrt{\frac{\kappa^{1/2}\log n }{nh^{1+2r}}}\right)$$
		as $h\rightarrow 0$, $\kappa \rightarrow \infty$ and $\frac{nh^{1+2r}}{\kappa^{1/2}\log n}\rightarrow \infty$. Moreover
		$$ \sup_{\theta,y}\bigg|\bigg|\frac{\partial^r}{\partial \theta^r}\hat{f}(\theta,y)-\frac{\partial^r}{\partial y^r}f(\theta,y)\bigg|\bigg|= O(\kappa^{-1}+h^2) + O_P\left(\sqrt{\frac{\kappa^{(1+2r)/2}\log n }{nh}}\right)$$
		as $h\rightarrow 0$, $\kappa \rightarrow \infty$ and $\frac{nh}{\kappa^{(1+2r)/2}\log n}\rightarrow \infty$. Finally,
		$$ \sup_{\theta,\phi}\bigg|\bigg|\frac{\partial^r}{\partial \phi^r}\hat{f}(\theta,\phi)-\frac{\partial^r}{\partial \phi^r}f(\theta,\phi)\bigg|\bigg|= O(\kappa^{-1}+\nu^{-1}) + O_P\left(\sqrt{\frac{\kappa^{(1+2r)/2}\nu^{1/2} \log n}{n}}\right)$$
		as $\kappa \rightarrow \infty$, $\nu \rightarrow \infty$  and $\frac{n}{\kappa^{(1+2r)/2 \nu^{1/2}}\log n}\rightarrow \infty$.
\end{lemma}
For the first assertion we apply the method of  \citet{Gine_Guillou_2002} and \citet{Einmahl_Mason2005} noting that the circular kernel is bounded. The proof of the second and third statements is analogous to the proof of Theorem 4 in \citet{Zhang_Chen2020} by using that the kernels associated to the predictor variables, $X$ or $\Theta$, are also bounded. Lemma~\ref{lem:uniform} is necessary for obtaining the consistency of the modal regression estimators since we need that the uniform errors tend to zero as the sample size increases.

\subsection{Consistency of the multimodal regression estimators}
Now we study the consistency of the multimodal regression estimators introduced in Section~\ref{sec:modal_reg_circ}. Note that usual error metrics in kernel regression (such as the Mean Integrated Squared Error) are not adequate to measure the quality of the estimators since in a regression multifunction we have, for each value of the predictor, a set of values of the response. Consequently, we will follow the work of \citet{Chen_etal2016} and consider pointwise and mean integrated errors based on a distance between sets.

Commence by considering the Hausdorff distance which, for two sets $A,B\subset \mathbb{R}^q$ is defined as
\begin{equation}
\mbox{Haus}(A,B)  =\max \bigg\{\sup_{x\in A } d(x,B), \sup_{x\in B}d(x,A)\bigg\},
\label{eq:haus}
\end{equation}
where, $d(x,A)=\inf_{z\in A}||x-z||$. The Hausdorff distance measures how close two Euclidean sets are from each other. In the real-valued response case exposed in Section~\ref{subsec:real-valued_response}, the Hausdorf distance is an adequate measure of the distance from the true regression multifunction to the estimated multifunction, because $M(\theta),\hat{M}(\theta)\subset \mathbb{R}$. However, in the circular response scenario our multifunctions are subsets of $\mathbb{T}$ and thus we need to generalize the definition of the Hausdorff distance by considering, for $A,B\subset \mathbb{T}$, 
\begin{equation}
\widetilde{\mbox{Haus}}(A,B)  =\max \bigg\{\sup_{x\in A } \tilde{d}(x,B), \sup_{x\in B}\tilde{d}(x,A)\bigg\},
\label{eq:haus_circ}
\end{equation}
with $\tilde{d}(x,A)=\inf_{z\in A} 1-\cos(x-z)$. The Hausdorff distances defined in (\ref{eq:haus}) and (\ref{eq:haus_circ}) are used to construct the pointwise error metrics, defined as 
\begin{eqnarray*}
	\Lambda(\theta)=\mbox{Haus}(M(\theta),\hat{M}(\theta)),\quad 
	\tilde{\Lambda}(\delta)=\widetilde{\mbox{Haus}}(M(\delta),\hat{M}(\delta)), 
	\label{eq:pointwise_error}
\end{eqnarray*}
where the first pointwise error is applied to the multifunctions with a circular predictor and real-valued response and the second is used for multifunctions with a circular response.
%The uniform error is defined as
%$$ \Lambda = \sup_{\delta \in \text{Supp}(\Delta)}\Lambda(\delta),$$
%with $\mbox{Supp}(\Delta)$ being the support of the variable $\Delta$ (a subset of $\mathbb{R}$ if $\Delta=X$ or $\mathbb{T}$ if $\Delta=\Theta$). 
The pointwise errors measure how close the estimated multifunction is from the true regression multi-valued function for each value of the predictor. In order to obtain a global error measure, we consider the modal Mean Integrated Squared Error ($\text{MISE}_m$) and the modal Circular Mean Integrated Error ($\text{CMIE}_m$), defined as

$$ \mbox{MISE}_m(\hat{M})=\mathbb{E}\left[\int_{\theta\in \mathbb{T}}\Lambda^2(\theta)d\theta\right], $$
$$\mbox{CMIE}_m(\hat{M})=\mathbb{E}\left[\int_{\delta\in \text{Supp}(\Delta)}\tilde{\Lambda}(\delta)d\delta\right], $$
where $\text{Supp}(\Delta)$ denotes the support of $\Delta$. The modal MISE is obtained by taking the expectation of the integrated squared pointwise error, and is a global error measure for the multimodal regression estimator in (\ref{eq:estimator_mutifunction_circlin}). The modal CMIE, constructed to measure the global error of the estimator in (\ref{eq:circ_resp_estim_Multifunction}), is the expectation of the integrated pointwise error. This error is in accordance with the integrated version of the so-called \textit{circular mean squared error} defined by \citet{Kim_SenGupta2017} as a circular analogous of the Mean Squared Error for euclidean variables.

With the objective of obtaining the asymptotic error rates for the multimodal estimators we must adopt the following assumptions on the conditional densities:
\begin{itemize}
	\item \textbf{(M1)} There exists $\lambda_1>0$ such that for any $(\theta,y)\in \mathbb{T}\times \Omega$ meeting $\frac{\partial}{\partial y}f(\theta,y)=0$, $\bigg|\frac{\partial^2}{\partial y^2}f(\theta,y)\bigg|<\lambda_1 $. In addition, for each $\theta\in \mathbb{T}$ there exist $B_1,\rho_1>0$ such that 
	$$\bigg\{y\in \Omega : \max_{y\in\Omega} \bigg|\frac{\partial}{\partial y}f(\theta,y)\bigg|\leq B_1, \frac{\partial^2}{\partial^2 y}f(\theta,y)\leq -\lambda_1<0\bigg\}\subset \mathcal{M}(\theta)\oplus\rho_1, $$
	where $\mathcal{M}(\theta)$ denotes the collection of local modes at $\Theta=\theta$ and $A\oplus r=\{x\in\mathbb{R}: \inf_{z\in A}|z-x|<r\}$.
	
	\item \textbf{(M2)} There exists $\lambda_2>0$ such that for any $(\delta,\phi)\in \text{Supp}(\Delta)\times \mathbb{T}$ meeting $\frac{\partial}{\partial \phi}f(\delta,\phi)=0$, $\bigg|\frac{\partial^2}{\partial \phi^2}f(\delta,\phi)\bigg|<\lambda_2 $. In addition, for each $\delta\in \text{Supp}(\Delta)$ there exist $B_2,\rho_2>0$ such that 
	$$\bigg\{\phi\in \mathbb{T} : \max_{\phi\in \mathbb{T}} \bigg|\frac{\partial}{\partial \phi}f(\delta,\phi)\bigg|\leq B_2, \frac{\partial^2}{\partial^2 \phi}f(\delta,\phi)\leq -\lambda_2<0\bigg\}\subset \mathcal{M}(\delta)\oplus\rho_2, $$
	where $\mathcal{M}(\delta)$ denotes the collection of local modes at $\Delta=\delta$ and $A\oplus r=\{\gamma\in\mathbb{T}: \inf_{z\in A} (1-\cos(z-\gamma))<r\}$.
	
\end{itemize} 
Conditions \textbf{(M1)} and \textbf{(M2)} guarantee that every conditional local mode is isolated from other critical values and  that points with near zero first partial derivative and negative second partial derivative are close to a conditional local mode.

\begin{prop}
	Let $\Theta$ be a circular random variable and $Y$ a real-valued random variable. Consider the multimodal regression estimator in (\ref{eq:estimator_mutifunction_circlin}). Under assumptions \textbf{(A1)}, \textbf{(A2)}, \textbf{(A4)}, \textbf{(K1)} and \textbf{(M1)} we have
	$$\Lambda(\theta)=O(\kappa^{-1}+h^2)+O_P\left(\sqrt{\frac{\kappa^{1/2}}{nh^3}}\right) $$
%	$$\Lambda=O(\kappa^{-1}+h^2)+O_P\left(\sqrt{\frac{\kappa^{1/2}\log n}{nh^3}}\right)$$
	and
	$$\mbox{MISE}_m(\hat{M})=O(\kappa^{-2}+h^4)+O_P\left(\frac{\kappa^{1/2}}{nh^3}\right). $$
\end{prop}

\begin{prop}
	Let $X$ be a real-valued  random variable and  $\Phi$ a  circular random variable. Consider the multimodal regression estimator in (\ref{eq:circ_resp_estim_Multifunction}). Under assumptions \textbf{(A1)}, \textbf{(A3)}, \textbf{(A4)}, \textbf{(K2)} and \textbf{(M2)}  we have
	$$\tilde{\Lambda}(x)=O(h^2+\kappa^{-1})+O_P\left(\sqrt{\frac{\kappa^{3/2}}{nh}}\right) $$
	%$$\Lambda=O(h^2+\kappa^{-1})+O_P\left(\sqrt{\frac{\kappa^{3/2}\log n}{nh}}\right)$$
	and
	$$\mbox{CMIE}_m(\hat{M})=O(h^2+\kappa^{-1})+O_P\left(\sqrt{\frac{\kappa^{3/2}}{nh}}\right). $$
\end{prop}

\begin{prop}
	Let $\Theta$ and $\Phi$ be circular random variables. Consider the multimodal regression estimator in (\ref{eq:circ_resp_estim_Multifunction}). Under assumptions \textbf{(A1)}, \textbf{(A3)}, \textbf{(A4)}, \textbf{(K2)} and \textbf{(M2)} we have
	$$\tilde{\Lambda}(\theta)=O(\nu^{-1}+\kappa^{-1})+O_P\left(\sqrt{\frac{\kappa^{3/2}\nu^{1/2}}{n}}\right) $$
%	$$\Lambda=O(\nu^{-1}+\kappa^{-1})+O_P\left(\sqrt{\frac{\kappa^{3/2} \nu^{1/2} \log n}{n}}\right)$$
	and
	$$\mbox{CMIE}_m(\hat{M})=O(\nu^{-1}+\kappa^{-1})+O_P\left(\sqrt{\frac{\kappa^{3/2}\nu^{1/2}}{n}}\right). $$
\end{prop}

\section{Bandwidth selection}
\label{sec:bandwidth}

As in most kernel methods, the selection of the smoothing parameters is a crucial issue in multimodal regression. In classical kernel methods for regression in the real line \citep{FanGijbels1996}, a large value of the bandwidth $h$ leads to an oversmoothed estimator, while a small value of $h$ produces an undersmoothed estimation of the regression function. Similarly to the context of circular kernel density estimation exposed in Section~\ref{subsec:real-valued_response}, the behavior of the concentration parameter in circular kernel regression \citep{DiMarzio_etal2009,DiMarzio_etal2012} is  reversed: when $\kappa$ is large, an undersmoothed estimation of the regression function is obtained, and a small value of $\kappa$ leads to an oversmoothed estimator. 

\begin{figure}[t]
	\centering
	\subfloat[$\kappa=5$]{
		\includegraphics[width=0.3\textwidth]{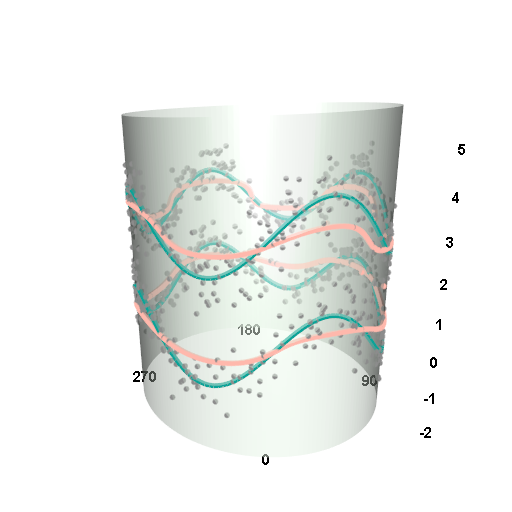}
		\label{fig:example_band1}}
	\hfill
	\subfloat[$\kappa=60$]{
		\includegraphics[width=0.3\textwidth]{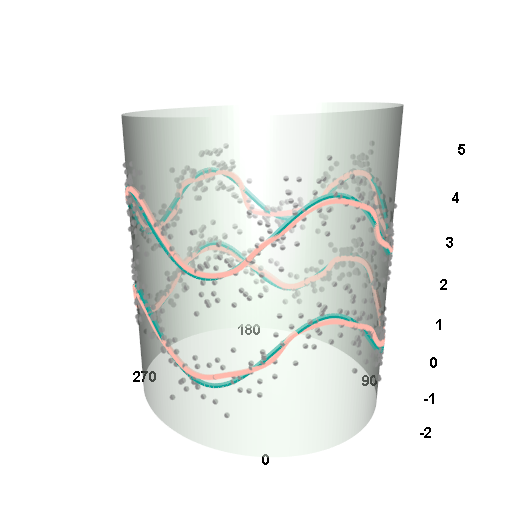}
		\label{fig:example_band2}}
	\hfill
	\subfloat[$\kappa=300$]{
		\includegraphics[width=0.3\textwidth]{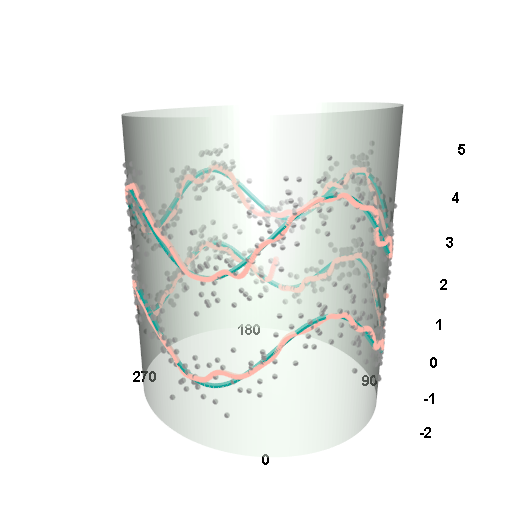}
		\label{fig:example_band3}} 
	
	\bigskip % some vertical space between subfigures a/b and c/d...
	\subfloat[$h=1.5$]{
		\includegraphics[width=0.3\textwidth]{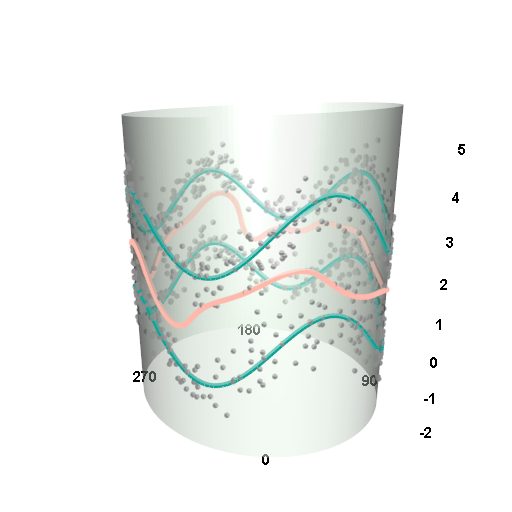}
		\label{fig:example_band4}}
	\hfill
	\subfloat[$h=0.5$]{
		\includegraphics[width=0.3\textwidth]{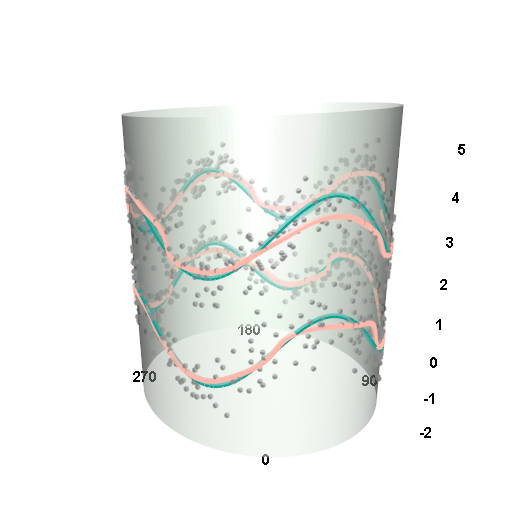}
		\label{fig:example_band5}}
	\hfill
	\subfloat[$h=0.2$]{
		\includegraphics[width=0.3\textwidth]{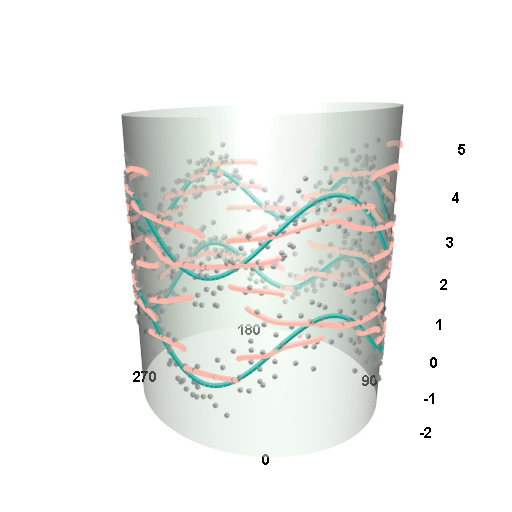}
		\label{fig:example_band6}} 
	
	\caption{Simulated data from a multimodal regression data with a circular predictor and a real-valued response. In green, the true regression multifunction and in pink the estimated multifunction. In the top row the smoothing parameter $h=0.6$ and the values of $\kappa$ are specified below each panel. In the bottom row the value of $\kappa$ is 30 and the values of $h$ are specified below each panel. }
	\label{fig:examples_bandwidth}
\end{figure}

However, in the context of multimodal regression two smoothing parameters are needed, one associated to the predictor variable and one associated to the response, and the role of the two parameters is very different. The parameter associated to the predictor variable controls the smoothing of the regression multifunction, having a similar role to the smoothing parameter in classical kernel regression. On the contrary, the parameter associated to the response variable affects the number of estimated modes. The reason is that an undersmoothed estimator of the conditional density will lead to many local modes, producing a large number of estimated branches in the regression multifunction. We illustrate this behavior in Figure~\ref{fig:examples_bandwidth} with simulated data from a multimodal regression model with a circular predictor and a real-valued response, although similar interpretations are obtained for the scenarios involving a circular response variable. In the top row of Figure~\ref{fig:examples_bandwidth} we fix the parameter $h$, associated to the response variable, and change $\kappa$, the parameter associated to the covariate. As the value of the concentratoin increases, the multimodal estimator turns more undersmoothed. On the other hand, in the bottom row the value of $\kappa$ was fixed and we change the value of $h$. If $h$ is too large, the two branches of the regression multifunction are merged into one, but if $h$ is too small, many local modes are estimated for each value of the predictor.

The statistical literature on the selection of the smoothing parameters for multimodal regression in the Euclidean setting is quite scarce. Since the estimation is carried out by obtaining the local maxima of the conditional density estimator, \citet{Einbeck_Tutz2006} recommend using methods for bandwidth selection tailored for conditional density estimation. However, such methods are not adequate in practice because, although related, the problem of mode estimation is not the same as the problem of density estimation. As noted by \citet{Casa_etal_2020}, an estimated density might be close to the true density in terms of ISE, but have many estimated local modes which in the case of modal regression would lead to many estimated branches of the regression multifunction. \citet{Zhou_Huang2019} propose two different methods for obtaining smoothing parameters in multimodal regression for real-valued variables. The first of them, called modal cross-validation, aims to balance the number of estimated local modes and the distance from the estimation to the data. Although the method shows a good performance in practice, nothing assures that minimizing the modal cross-validation function will minimize the modal MISE of the estimator. The second approach in \citet{Zhou_Huang2019} is to minimize the modal ISE of the estimator (the integrated Hausdorff distance between the estimator and the true multifunction) by using a parametric bootstrap in which the parametric model is a mixture of regressions. Another approach is given by \citet{Chen_etal2016}, which consists on constructing a prediction band for the regression multifunction and afterwards selecting the parameters minimizing a loss function defined as the volume of the prediction band. However, this method has several drawbacks. First, the authors assume that the smoothing parameter is the same for both variables, which does not seem realistic, especially given the different role of the two parameters. In addition, when two different parameters are used, the approach tends to select values which produce many estimated local modes.

\subsection{Modal cross-validation for circular regression}

Although the theoretical properties of the modal cross-validation method in \citet{Zhou_Huang2019} are not well studied, the behavior of the method in practice is quite satisfactory. One of the advantages of the modal cross-validation method is that it is almost immediately adaptable to the context of circular multimodal regression. For the scenario presented in Section~\ref{subsec:real-valued_response} (circular predictor and real-valued response), the modal CV approach consists on selecting $\kappa$ and $h$ minimizing
$$ CV_1(\kappa,h)=\frac{1}{n}\sum_{i=1}^{n}d^2(\hat{M}_{-i}(\Theta_i),Y_i)N^2_{-i}(\Theta_i), $$
where $d(x,A)=\inf_{z\in A}||x-z||$, $\hat{M}_{-i}$ is the estimator of $M$ using data $\{(\Theta_j,Y_j):j\neq i\}$ and $N_{-i}(\Theta_i)$ is the number of estimated modes of $\hat{M}_{-i}$ at $\theta=\Theta_i$. For the circular response case exposed in Section~\ref{subsec:circular_response}, the parameters $g$ and $\kappa$ (where $g$ represents either $h$ or $\nu$ depending on the nature of the predictor) are selected by minimizing 
$$ CV_2(g,\kappa)=\frac{1}{n}\sum_{i=1}^{n}\tilde{d}(\hat{M}_i(\Delta_i),Y_i)N^2_{-i}(\Delta_i), $$
where $\tilde{d}(x,A)=\inf_{z\in A} 1-\cos(x-z)$.
This approach does not lie on a theoretical basis and computational trials are needed in order to assess the efficacy of the method.

\subsection{Parametric bootstrap for circular regression}
The parametric bootstrap approach in \citet{Zhou_Huang2019} can also be adapted for circular data. We start by adapting the method for circular predictors and real-valued responses, considering the multimodal estimator in (\ref{eq:estimator_mutifunction_circlin}). The goal of this method is to minimize the modal ISE, defined as
\begin{equation}
	 \mbox{ISE}_m(\hat{M})=\int_{\theta  \in \mathbb{T}} \Lambda^2(\theta) d\theta,
	 \label{eq:ISE}
\end{equation}
which is the integrated version of the squared pointwise error. In order to estimate (\ref{eq:ISE}), we construct an empirical version
$$ \frac{1}{n}\sum_{j=1}^{n}\Lambda^2(\Theta_j)=\frac{1}{n}\sum_{j=1}^{n}\mbox{Haus}^2(\hat{M}(\Theta_j),M(\Theta_j)), $$
which depends on the true regression multifunction. We estimate the conditional density $f(y|\theta)$ with a mixture of parametric conditional densities, obtaining $\tilde{f}(y|\theta)$. Bootstrap resamples of the data $\{(\Theta_j,Y_j^{*(b)})\}_{j=1}^n$, with $b=1,...,B$ are drawn using $\tilde{f}(y|\theta)$, where $B$ is the number of boostrap resamples. Thus, the 
empirical version of (\ref{eq:ISE}) is estimated as 
$$\frac{1}{B}\sum_{b=1}^{B}\frac{1}{n}\sum_{j=1}^{n}\mbox{Haus}^2(\hat{M}^{*(b)}(\Delta_j),\tilde{M}(\Delta_j)), $$
where $\tilde{M}$ is a mode estimator of $\tilde{f}(y|\theta)$ and $\hat{M}^{*(b)}$ is the multimodal estimator of $M$ based on the $b$th resample of the data. For the parametric estimator  $\tilde{f}(y|\theta)$, we use a mixture of normal densities 
$$ \tilde{f}(y|\theta)=\sum_{t=1}^T\pi_t \mathcal{N}(y-\beta_{0t}-\beta_{1t}b_1(\theta)-...-\beta_{kt}b_k(\theta),\sigma^2), $$
where $\pi_t$, $t=1,..,T$ are non-negative weights summing up to one, $\{b_1,...,b_k\}$ is a basis o periodic B-Splines to account for the circular nature of $\Theta$ and the parameters $\beta_{0t},...,\beta_{kt}, \sigma^2$ are estimated from the data. The number of components in the mixture, $T$, is selected by the BIC criterion.

Consider now the context of estimating the regression multifunction for a circular response with the estimator in (\ref{eq:circ_resp_estim_Multifunction}). We want to minimize the modal Circular Integrated Error, defined as
\begin{equation*}
\mbox{CIE}_m(\hat{M})=\int_{\delta  \in \text{Supp}(\Delta)}\tilde{\Lambda}(\delta) d\delta.
\label{eq:CIE}
\end{equation*}
or, in practice, its empirical version
$$ \frac{1}{n}\sum_{j=1}^{n}\tilde{\Lambda}(\Delta_j)=\frac{1}{n}\sum_{j=1}^{n}\widetilde{\mbox{Haus}}(\hat{M}(\Delta_j),M(\Delta_j)). $$
In order to estimate the empirical version of $\mbox{CIE}_m(\hat{M})$, we use the same approach as above, by estimating the conditional density parametrically and then using such estimator to construct bootstrap resamples. The only difference lies on the parametric estimator of the conditional density. In the case where the predictor is real-valued, we use a mixture of regressions of projected normals with means depending on the predictor \citep{Presnell_etal1998,Maruotti_2016}. For the case where both variables have a circular nature, we may consider a mixture of the models in \citet{Kim_SenGupta2017}.

%\section{Discussion}
%\label{sec:conclussions}

\section{Acknowledgements}
The authors acknowledge the financial support of Project MTM2016-76969-P from the AEI co-funded by the European Regional Development Fund (ERDF) and the Competitive Reference Groups 2017-2020 (ED431C 2017/38) from the Xunta de Galicia through the ERDF. Research of Mar\'ia Alonso-Pena was supported by the Xunta de Galicia grant ED481A-2019/139 through \textit{Programa de axudas \'a etapa predoutoral da Xunta de Galicia (Conseller\'ia de Educaci\'on, Universidade e Formaci\'on Profesional)}. 

% Bibliography

\phantomsection

\bibliographystyle{apalike}
\bibliography{biblio_modal_reg_circ}

\begin{thebibliography}{}

\bibitem[Aliyari~Ghassabeh, 2013]{AliyariGhassabeh_2015}
Aliyari~Ghassabeh, Y. (2013).
\newblock A sufficient condition for the convergence of the mean shift
  algorithm with gaussian kernel.
\newblock {\em Pattern Recognition Letters}, 34(12):1423--1427.

\bibitem[Aliyari~Ghassabeh, 2015]{AliyariGhassabeh_2013}
Aliyari~Ghassabeh, Y. (2015).
\newblock On the convergence of the mean shift algorithm in the one-dimensional
  space.
\newblock {\em Journal of Multivariate Analysis}, 135(1):1--10.

\bibitem[Aliyari~Ghassabeh et~al., 2013]{AliyariGhassabeh_etal2013}
Aliyari~Ghassabeh, Y., Linder, T., and Takahara, G. (2013).
\newblock On some convergence properties of the subspace constrained mean
  shift.
\newblock {\em Pattern Recognition}, 46(11):3140--3147.

\bibitem[Ameijeiras-Alonso et~al., 2019]{Ameijeiras-Alonso_etal2019b}
Ameijeiras-Alonso, J., Lagona, F., Ranalli, M., and Crujeiras, R. (2019).
\newblock A circular nonhomogeneous hidden markov field for the spatial
  segmentation of wild fire occurrences.
\newblock {\em Environmetrics}, 30(2).

\bibitem[Bai et~al., 1988]{Bai_etal1988}
Bai, Z., Rao, C., and Zhao, L. (1988).
\newblock Kernel estimators of density function of directional data.
\newblock {\em Journal of Multivariate Analysis}, 27(1):24--39.

\bibitem[Bonnabel, 2013]{Bonnabel_2015}
Bonnabel, S. (2013).
\newblock Stochastic gradient descent on riemannian manifolds.
\newblock {\em IEEE Transactions on Automatic Control}, 58(9):2217--2229.

\bibitem[Casa et~al., 2020]{Casa_etal_2020}
Casa, A., Chac\'on, J., and Menardi, G. (2020).
\newblock Modal clustering asymptotics with applications to bandwidth
  selection.
\newblock {\em Electronic Journal of Statist}, 14(1):835--856.

\bibitem[Chen, 2018]{Chen2018}
Chen, Y.-C. (2018).
\newblock Modal regression using kernel density estimation: A review.
\newblock {\em WIREs Computational Statistics}, 10:e1431.

\bibitem[Chen et~al., 2014]{Chen_etal2014}
Chen, Y.-C., Genovese, C., , and Wasserman, L. (2014).
\newblock {Generalized Mode and Ridge Estimation}.
\newblock {\em arXiv e-prints}, page arXiv:1406.1803.

\bibitem[Chen et~al., 2016]{Chen_etal2016}
Chen, Y.-C., Genovese, C., Tibshirani, R., and Wasserman, L. (2016).
\newblock Nonparametric modal regression.
\newblock {\em Annals of Statistics}, 44(2):489--514.

\bibitem[Cheng, 1995]{Cheng1995}
Cheng, Y. (1995).
\newblock Mean shift, mode seeking, and clustering.
\newblock {\em IEEE Transactions on Pattern Analysis and Machine Intelligence},
  17(8):790--799.

\bibitem[Comaniciu and Meer, 2002]{Comaniciu_Meer2002}
Comaniciu, D. and Meer, P. (2002).
\newblock Mean shift: a robust approach toward feature space analysis.
\newblock {\em IEEE Trans Pattern Anal Mach Intell}, 24(5):603--619.

\bibitem[Di~Marzio et~al., 2016]{DiMarzio_etal2016}
Di~Marzio, M., Fensore, S., Panzera, A., and Taylor, C. (2016).
\newblock A note on nonparametric estimation of circular conditional densities.
\newblock {\em Journal of Statistical Computation and Simulation},
  86(13):2573--2582.

\bibitem[Di~Marzio et~al., 2009]{DiMarzio_etal2009}
Di~Marzio, M., Panzera, A., and Taylor, C. (2009).
\newblock Local polynomial regression for circular predictors.
\newblock {\em Statistics \& Probability Letters}, 798(1):2066--2075.

\bibitem[Di~Marzio et~al., 2011]{DiMarzio_etal2011}
Di~Marzio, M., Panzera, A., and Taylor, C. (2011).
\newblock Kernel density estimation on the torus.
\newblock {\em Journal of Statistical Planning and Inference},
  141(1):2156--2173.

\bibitem[Di~Marzio et~al., 2012]{DiMarzio_etal2012}
Di~Marzio, M., Panzera, A., and Taylor, C. (2012).
\newblock Non-parametric regression for circular responses.
\newblock {\em Scandinavian Journal of Statistics}, 40(2):238--255.

\bibitem[Einbeck and Tutz, 2006]{Einbeck_Tutz2006}
Einbeck, J. and Tutz, G. (2006).
\newblock Modelling beyond regression functions: An application of multimodal
  regression to speedflow data.
\newblock {\em Journal of the Royal Statistical Society Series C (Applied
  Statistics)}, 55(4):461--475.

\bibitem[Einmahl and Mason, 2005]{Einmahl_Mason2005}
Einmahl, U. and Mason, D. (2005).
\newblock Uniform in bandwidth consistency of kernel-type function estimators.
\newblock {\em Annals of Statistics}, 33(3):1380--1403.

\bibitem[Fan and Gijbels, 1996]{FanGijbels1996}
Fan, J. and Gijbels, I. (1996).
\newblock {\em Local Polynomial Modelling and its Applications}.
\newblock Chapman and Hall, London.

\bibitem[Fisher, 1989]{Fisher1989}
Fisher, N. (1989).
\newblock Smoothing a sample of circular data.
\newblock {\em Journal of Structural Geology}, 11(6):775--778.

\bibitem[Fukunaga and Hostetler, 1975]{Fukunaga_Hostetler1975}
Fukunaga, K. and Hostetler, L. (1975).
\newblock The estimation of the gradient of a density function, with
  applications in pattern recognition.
\newblock {\em IEEE Transactions on Information Theory}, 21(1):32--40.

\bibitem[Garc\'ia-Portugu\'es et~al., 2013]{GarciaPortugues_etal2013}
Garc\'ia-Portugu\'es, E., Crujeiras, R., and Gonz\'alez-Manteiga, W. (2013).
\newblock Kernel density estimation for directional-linear data.
\newblock {\em Journal of Multivariate Analysis}, 121(1):152--275.

\bibitem[Gine and Guillou, 2002]{Gine_Guillou_2002}
Gine, E. and Guillou, A. (2002).
\newblock Rates of strong uniform consistency for multivariate kernel density
  estimators.
\newblock {\em Annales de l'Institut Henri Poincare (B) Probability and
  Statistics}, 38(6):907--921.

\bibitem[Jammalamadaka and SenGupta, 2001]{Jammalamadaka_SenGupta_2001}
Jammalamadaka, S. and SenGupta, A. (2001).
\newblock {\em Topics in Circular Statistics.}
\newblock World Scientific, Singapore.

\bibitem[Jander, 1957]{Jander1957}
Jander, R. (1957).
\newblock Die optische richtungsorientierung der roten waldameise (formicaruea
  l.).
\newblock {\em Zeitschrift für vergleichende Physiologie}, 40(1):162--238.

\bibitem[Jona-Lasinio et~al., 2012]{Jona-Lasinio_etal2012}
Jona-Lasinio, G., Gelfand, A., and Jona-Lasinio, M. (2012).
\newblock Spatial analysis of wave direction data using wrapped gaussian
  processes.
\newblock {\em Annals of Applied Statistics}, 6(4):1478--1498.

\bibitem[Kim and SenGupta, 2017]{Kim_SenGupta2017}
Kim, S. and SenGupta, A. (2017).
\newblock Multivariate-multiple circular regression.
\newblock {\em Journal of Statistical Computation and Simulation},
  87(7):1277--1291.

\bibitem[Kobayashi and Otsu, 2010]{Kobayashi_Otsu_2010}
Kobayashi, T. and Otsu, N. (2010).
\newblock Von mises-fisher mean shift for clustering on a hypersphere.
\newblock In {\em Proceedings of the 20th international conference on pattern
  recognition}, pages 2130--2133. IEEE.

\bibitem[Ley and Verdebout, 2017]{Ley_Verdebout2017}
Ley, C. and Verdebout, T. (2017).
\newblock {\em Modern Directional Statistics.}
\newblock Chapman \& Hall, Boca Raton.

\bibitem[Li et~al., 2007]{Li_etal2007}
Li, X., Hu, Z., and Wu, F. (2007).
\newblock A note on the convergence of the mean shift.
\newblock {\em Pattern Recognition}, 40(6):1756--1762.

\bibitem[Mardia and Jupp, 2000]{Mardia_Jupp_2000}
Mardia, K. and Jupp, P. (2000).
\newblock {\em Directional Statistics.}
\newblock John Wiley \& Sons, Inc., New York.

\bibitem[Maruotti, 2016]{Maruotti_2016}
Maruotti, A. (2016).
\newblock Analyzing longitudinal circular data by projected normal models: a
  semi-parametric approach based on finite mixture models.
\newblock {\em Environmental and Ecological Statistics}, 23(1):257–277.

\bibitem[Mooney et~al., 2003]{Mooney_etal2003}
Mooney, J., Helms, P., and Jollife, I. (2003).
\newblock Fitting mixtures of von mises distributions: a case study involving
  sudden infant death syndrome.
\newblock {\em Computational Statistics \& Data Analysis}, 41(1):505--513.

\bibitem[Oba et~al., 2005]{Oba_etal2005}
Oba, S., Kato, K., and Ishii, S. (2005).
\newblock Multi-scale clustering for gene expression profiling data.
\newblock In {\em 5th IEEE Symposium on Bioinformatics and Bioengineering
  (BIBE'05)}, pages 210--217.

\bibitem[Oliveira et~al., 2014]{Oliveira_etal_2014}
Oliveira, M., Crujeiras, R., and Rodr\'iguez-Casal, A. (2014).
\newblock Circsizer: an exploratory tool for circular data.
\newblock {\em Journal of Environmental and Ecological Statistics},
  21(1):143--159.

\bibitem[Parzen, 1962]{Parzen1962}
Parzen, E. (1962).
\newblock On estimation of a probability density function and mode.
\newblock {\em Annals of Mathematical Statistics}, 33(1):1065--1076.

\bibitem[Pewsey et~al., 2013]{Pewsey_etal2013}
Pewsey, A., Neuhuser, M., and Ruxton, G. (2013).
\newblock {\em Circular Statistics in R.}
\newblock Oxford University Press, Oxford.

\bibitem[Presnell et~al., 1998]{Presnell_etal1998}
Presnell, B., Morrison, S., and Little, R. (1998).
\newblock Multivariate linear models for directional data.
\newblock {\em Journal of the American Statistical Association,},
  93(443):1068--1077.

\bibitem[Rosenblatt, 1956]{Rosenblatt1956}
Rosenblatt, M. (1956).
\newblock Remarks on some nonparametric estimate of a density function.
\newblock {\em Annals of Mathematical Statistics}, 27(1):832--837.

\bibitem[Scott, 1992]{Scott1992}
Scott, D. (1992).
\newblock {\em Multivariate Density Estimation}.
\newblock Wiley, New York.

\bibitem[SenGupta and Rao, 1966]{SenGupta_Rao1966}
SenGupta, S. and Rao, J. (1966).
\newblock Statistical analysis of cross-bedding azimuths from the kamthi
  formation around bheemaram, pranhita: Godavari valley.
\newblock {\em Sankhya: The Indian Journal of Statistics, Series B.},
  28(1):165--174.

\bibitem[Wand and Jones, 1995]{WandJones1995}
Wand, M. and Jones, M. (1995).
\newblock {\em Kernel Smoothing}.
\newblock Chapman \& Hall, London.

\bibitem[Zhang and Sra, 2016]{Zhang_Sra2016}
Zhang, H. and Sra, S. (2016).
\newblock First-order methods for geodesically convex optimization.
\newblock In {\em Proceedings of Machine Learning Reasearch}, pages 1617--1638.
  PMLR.

\bibitem[Zhang and Chen, 2020]{Zhang_Chen2020}
Zhang, Y. and Chen, Y.-C. (2020).
\newblock {Kernel smoothing, mean shift, and their learning theory with
  directional data}.
\newblock {\em arXiv e-prints}, page arXiv:2010.13523.

\bibitem[Zhou and Huang, 2019]{Zhou_Huang2019}
Zhou, H. and Huang, X. (2019).
\newblock Bandwidth selection for nonparametric modal regression.
\newblock {\em Communications in Statistics - Simulation and Computation},
  48(4):968--984.

\end{thebibliography}

\end{document}